\documentclass[twocolumn,epsfig,APS,floats,showpacs]{revtex4}
\usepackage{graphicx}
\usepackage{color}

\usepackage{fancyhdr}
\fancypagestyle{titlepage}{
\fancyhead[R]{J.  Magn. Magn. Phen. \textbf{422}, 362 (2016)}}
\begin{document}

\title{Induced magnetism at the interfaces of a Fe/V superlattice investigated by resonant magnetic x-ray scattering}

\author{Martin Magnuson}

\affiliation{Department of Physics, IFM, Thin Film Physics Division, Link\"{o}ping University, 
SE-58183 Link\"{o}ping, Sweden.}

\date{\today}

\begin{abstract}
The induced magnetic moments in the V $3d$ electronic states of interface atomic layers in a Fe(6ML)/V(7ML) superlattice was investigated by x-ray resonant magnetic scattering. The first V atomic layer next to Fe was found to be strongly antiferromagnetically polarized relatively to Fe and the magnetic moments of the next few atomic layers in the interior V region decay exponentially with increasing distance from the interface, while the magnetic moments of the Fe atomic layers largely remain bulk-like. The induced V moments decay more rapidly as observed by x-ray magnetic scattering than in standard x-ray magnetic circular dichroism. The theoretical description of the induced magnetic atomic layer profile in V was found to strongly rely on the interface roughness within the superlattice period. These results provide new insight into interface magnetism by taking advantage of the enhanced depth sensitivity to the magnetic profile over a certain resonant energy bandwidth in the vicinity of the Bragg angles.
\end{abstract}

\maketitle

\section{introduction}


In the quest for improved applications in magnetism, the fundamental mechanism for the magnetic coupling in superlattices for magnetic data storage and magnetic anisotropy has been in focus for a long time \cite{Book1}. Superlattices are also of technological interest as mirrors in the soft x-ray region using alternating layers of high and low density elements to significantly enhance the reflectivity in a narrow range around the Bragg angles \cite{Book2}. 
Metallic ferromagnetic (FM) layers such as Fe, that are separated by a nonmagnetic spacer layer such as V exhibit an oscillating interlayer exchange coupling (IEC) that is either FM or antiferromagnetic (AFM) depending on the spacer layer thickness \cite{Fert,Grunberg}. As a model system, Fe/V superlattices have been used to study the IEC \cite{Skubic}, the induced magnetic moments in V \cite{Scherz1}, and the giant magnetoresistance (GMR) effect \cite{Granberg}. Fe/V superlattices have also been shown to be possible hydrogen storage media whereby the hydrogen modify the electronic structure of the nonmagnetic (NM) spacer layer as well as the magnetic interlayer coupling \cite{Palsson}. When a Fe/V multilayer is grown in the (110) plane on a MgO(001) single crystal, the IEC between successive Fe layers oscillatory couple FM and AFM \cite{Broddefalk}. However, magnetometry measurements indicated a periodic oscillating coupling as a function of V interlayer thickness with Fe AFM coupled at 22 \AA{}, 32 \AA{} and 42 \AA{} \cite{Schwickert} but no AFM coupling peak was observed at 12 \AA{} (7 ML) V layer thickness. 

Previous studies of the size and extent of the V polarization in Fe/V superlattices have mainly been studied by x-ray magnetic circular dichroism (XMCD) at the $2p$ absorption edges in total electron yield (TEY) mode \cite{Stohr}. For Fe/V superlattices, it was found that the induced average V magnetic moments increase with thinner V or Fe layers \cite{Harp}. Although vanadium is non-magnetic in bulk form, it obtains a weakly induced magnetic moment in a Fe/V superlattice due to hybridization with Fe at the interfaces that is normally aligned antiparallel to those of Fe \cite{Scherz2}. 
More detailed XMCD studies came to the conclusion that the atomic layer resolved induced magnetic moments of V decay monotonically and slowly with distance from the Fe interface so that even the fourth and fifths atomic layer from the interface possessed a significant magnetic moment \cite{Tomaz,Clavero}. These observations are at variance with calculated results \cite{Niklasson}, where the induced magnetic moment of V is most significant at the first layer in contact with Fe while the magnetic moments in the interior atomic layers are negligible. However, the limited probe depth in the TEY mode of about 15 \AA{} in XMCD\cite{Magnuson1}, is not useful when capping layers are utilized to prevent surface oxidation of superlattices.

For the investigation of local magnetic properties in deep buried layers, bulk-sensitive x-ray magnetic scattering (XRMS) is better suited \cite{Tonnerre2}. 
While the wavelengths in the soft x-ray regime are usually too long for Bragg diffraction of single crystals, they are very suitable for larger periodic structures such as multilayers with lattice spacing of a few monolayers. With the use of circularly or elliptically polarized synchrotron radiation in the excitation, a dichroic XRMS spectrum is the difference in scattered intensity obtained with opposite relative orientations of the photon spin (helicity) of the incident x-rays and the applied magnetization direction of the sample \cite{Gibbs1,Gibbs2,Isaacs,McWhan,Hannon}. In previous XRMS investigations on thin superlattices, it was found that the technique is sensitive to changes in the optical constants \cite{Magnuson3} along the surface normal that depend on the local magnetic properties and therefore it can be used to distinguish between different shapes of magnetization depth profiles \cite{Seve,Sacchi1}. 

In this paper, we investigate the magnetic coupling and quantify the induced V magnetic moments at the interfaces of a (Fe 6 ML)/(V 7 ML) superlattice by taking advantage of the large probe depth and element selectivity of the XRMS technique at resonant conditions. By utilizing the enhanced magnetic sensitivity by the rapidly changing refractive index at energies around the $2p$ absorption resonances in combination with the constructive interference scattering at the interfaces in the vicinity of the Bragg angles, it is shown that it is possible to use a relatively limited data set to distinguish between different magnetization profiles of the weakly induced moments at the interface and in the interior regions of the V spacer layers of a Fe/V superlattice.

\begin{figure}
\includegraphics[width=85mm]{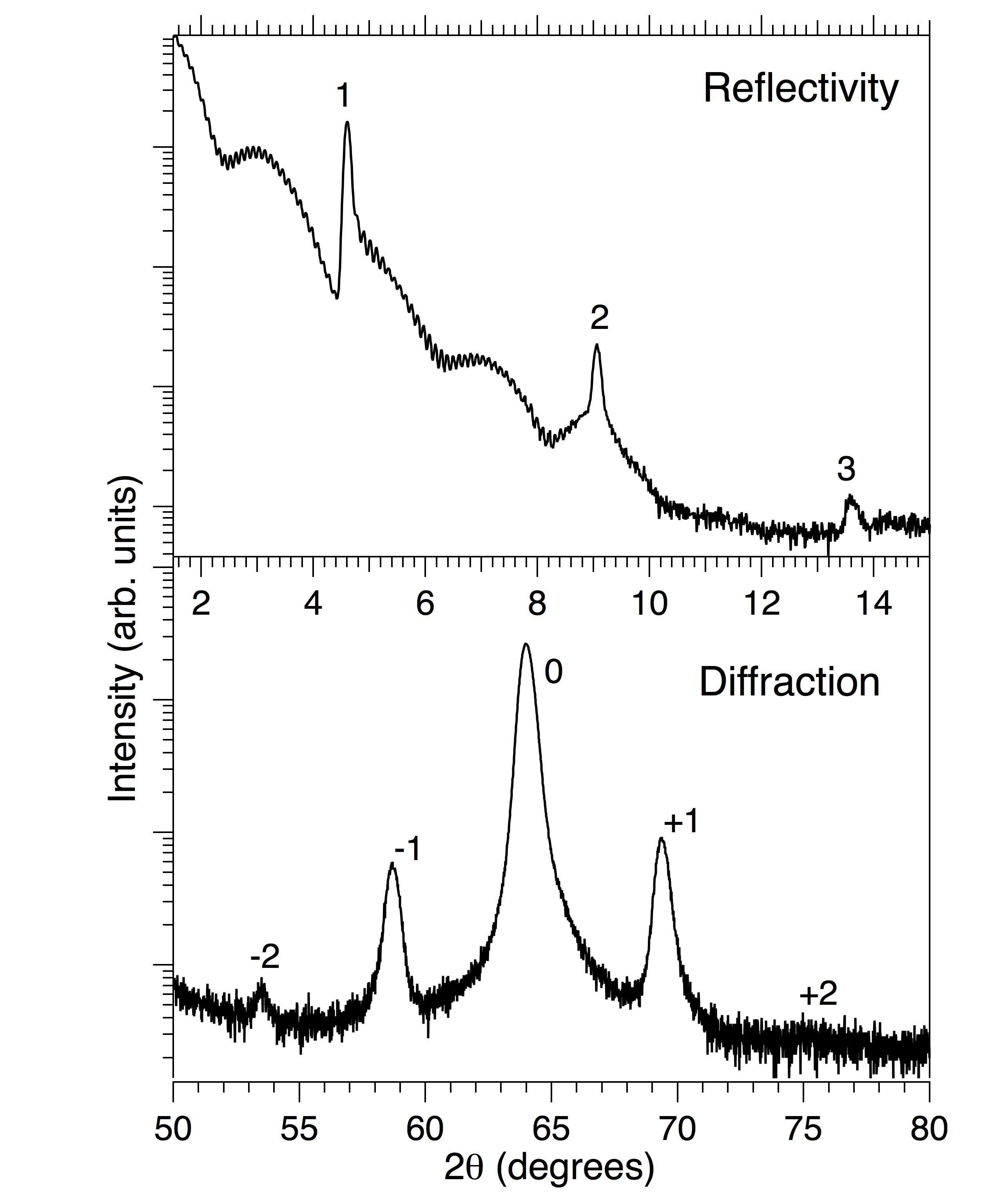}
\vspace{2ex}
\caption{X-ray reflectivity and diffraction of the Fe(6ML)/V(7ML) superlattice measured with conventional $\theta$-2$\theta$ Cu $K_{\alpha}$ radiation ($\lambda$=1.54  \AA{}) x-ray source. Top panel: Low-angle x-ray reflectivity data with the main x-ray diffraction (XRD) peaks indicated. b) Bottom panel: High-angle XRD where the main superlattice 002 peak indicated by "0" is surrounded by satellites denoted $\pm$1 and $\pm$2.}
\label{XRD}
\end{figure}

\section{experimental details}

The XRMS measurements were performed using the reflectometer at the soft x-ray metrology bending magnet beamline 6.3.2 at the Advanced Light Source (ALS) in Berkeley, USA \cite{Underwood1,Underwood2}.  The out-of-plane elliptically polarized radiation was extracted by using a four-jaw aperture in the beamline. The monochromator was set to a resolving power of 2000 and a flux of $10^{10}$ photons per second on the sample at the Fe and V $L_{3}$-edges. The sample was magnetized along the (100) easy axis, parallel to its surface and in the scattering plane (longitudinal mode), by means of a permanent magnet (0.1 T), situated behind the sample holder, mounted on a stepper motor, used to reverse the field direction at each photon energy. The energy scans were performed at three different $\theta$-angles at both the Fe and V $2p$ thresholds (680-730 eV and 500-545 eV, respectively) with a 0.2 eV step size. The incoming flux was monitored and used to normalize the spectra. The incident photons were 60 \% circularly polarized and the sample was mounted in the reflectometer end station with the axis of rotation parallel to the orbit plane. 

The single-crystal Fe/V thin film superlattice was epitaxially grown in ultrahigh vacuum by dual-target magnetron sputtering deposition of metallic Fe and V layers on a polished MgO(001) fcc single crystal substrate at 300$^{\circ}$C \cite{Isberg}. The superlattice was grown in the (110) plane and is thus rotated 45$^{\circ}$ with respect to the (100) direction of the substrate. A biaxial compressive strain on V and a tensile strain on Fe results in a body centered tetragonal (bct) structure due to the lattice strain and mismatch at the interface of 5.1 \% between bulk Fe and bulk V. The alternating depositions of the Fe and V layers were repeated to form a total of 40 bilayer periods and capped with a 40 \AA{} Pd film to prevent oxidation. 

Analysis of the structural parameters was made by x-ray diffraction (XRD) before fitting the spectroscopic XRMS part. The structural quality of the sample was checked and the layer thicknesses determined by using conventional $\theta$-2$\theta$ XRD measurements with Cu K$_{\alpha}$ radiation ($\lambda$=1.54  \AA{}) for low angles (1.5-15$^{\circ}$ in 2$\theta$) and high angles (50-80$^{\circ}$ in 2$\theta$) around the Fe/V (002) Bragg peak. 
Figure 1 shows reflectivity (top panel) and x-ray diffraction data (bottom panel) of the Fe/V superlattice. The low-angle data (top panel), show well-defined Bragg peaks denoted 1-3, arising from the chemical modulation, surrounded by small Kiessig fringes that appear from interference between the surface and bottom of the whole film. The appearance of the third Bragg peak in the reflectivity data indicates that the interface roughness is small. By fitting the angular positions of the Kiessig fringes to a linearization of BraggÕs law, the total film thickness was determined to be 830$\pm$2 \AA{} and the number of periods to 40.

From the high-angle diffraction measurements (bottom panel), the periodicity $\Lambda$ of the multilayer was determined from the angular positions of the intense (002) Bragg peak and the satellites according to Braggs law: $\Lambda$=n$\times$$\lambda$/[2(sin$\theta_{n}$-sin$\theta_{0}$)], where "0" is the main Bragg peak and n=$\pm1$ and $\pm$2 are the satellites on both sides. Table I lists the structural parameters of the individual layers with error bars obtained by a successive refinement procedure to reproduce the Bragg peaks of the XRD data using the computer program SUPREX \cite{Fullerton}. The periodicity $\Lambda$=$t_{1}$+$t_{2}$ was determined to be 19.7$\pm$0.1 \AA{} and the individual thicknesses of Fe 7.4$\pm$0.1 \AA{} (6ML) and V 12.3$\pm$0.1 \AA{} (7ML). From the XRD fitting, the Pd capping layer was determined to be 40$\pm$1 \AA{} thick.

\begin{figure}
\includegraphics[width=85mm]{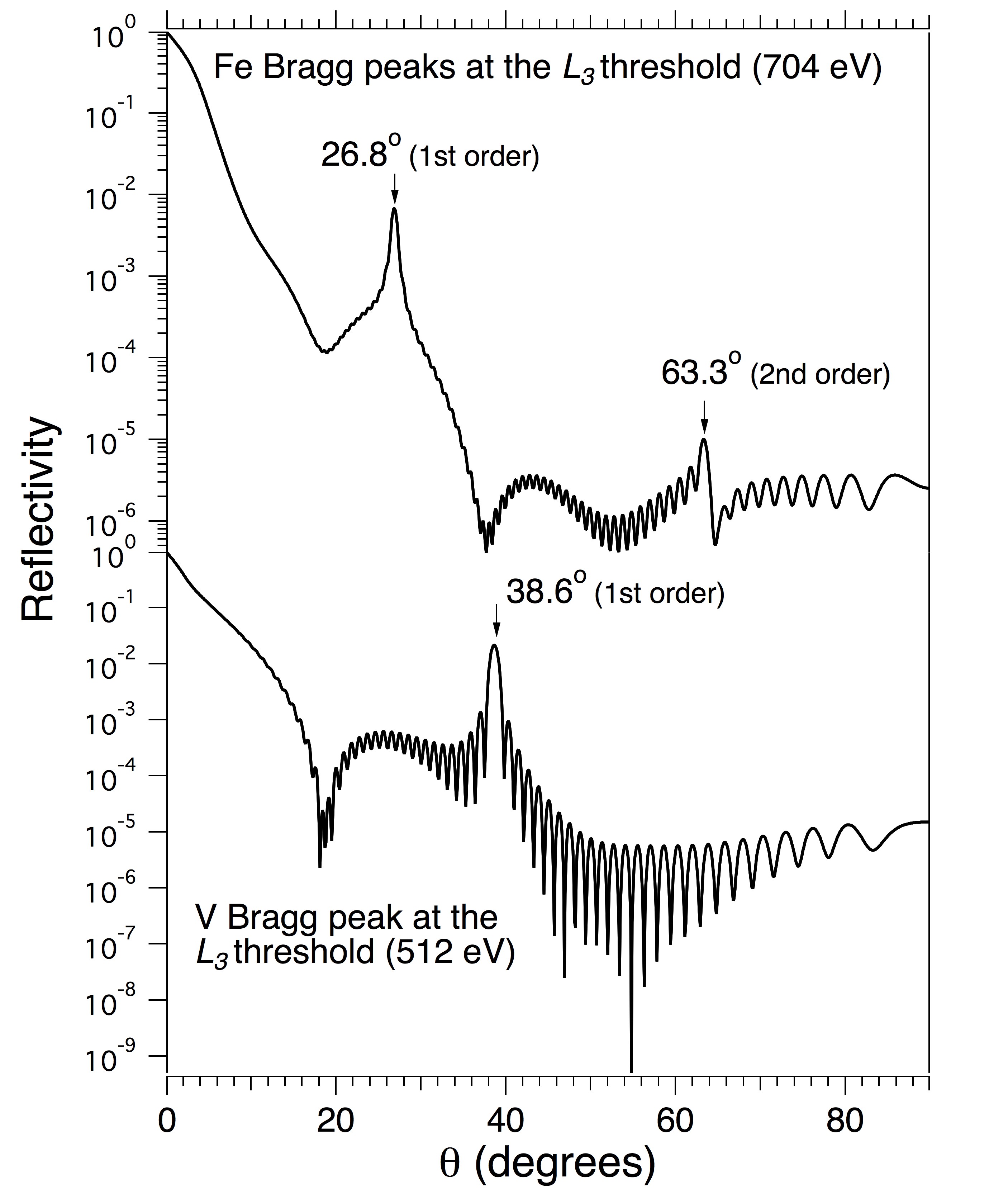}
\vspace{2ex}
\caption{Magnetization averaged x-ray reflectivity and diffraction measurements of the Fe(6ML)/V(7ML) superlattice with photon 
energies of 704 eV and 512 eV at the Fe and V $2p$ thresholds, respectively.}
\label{XRD_synchr}
\end{figure} 

\begin{figure}
\includegraphics[width=90mm]{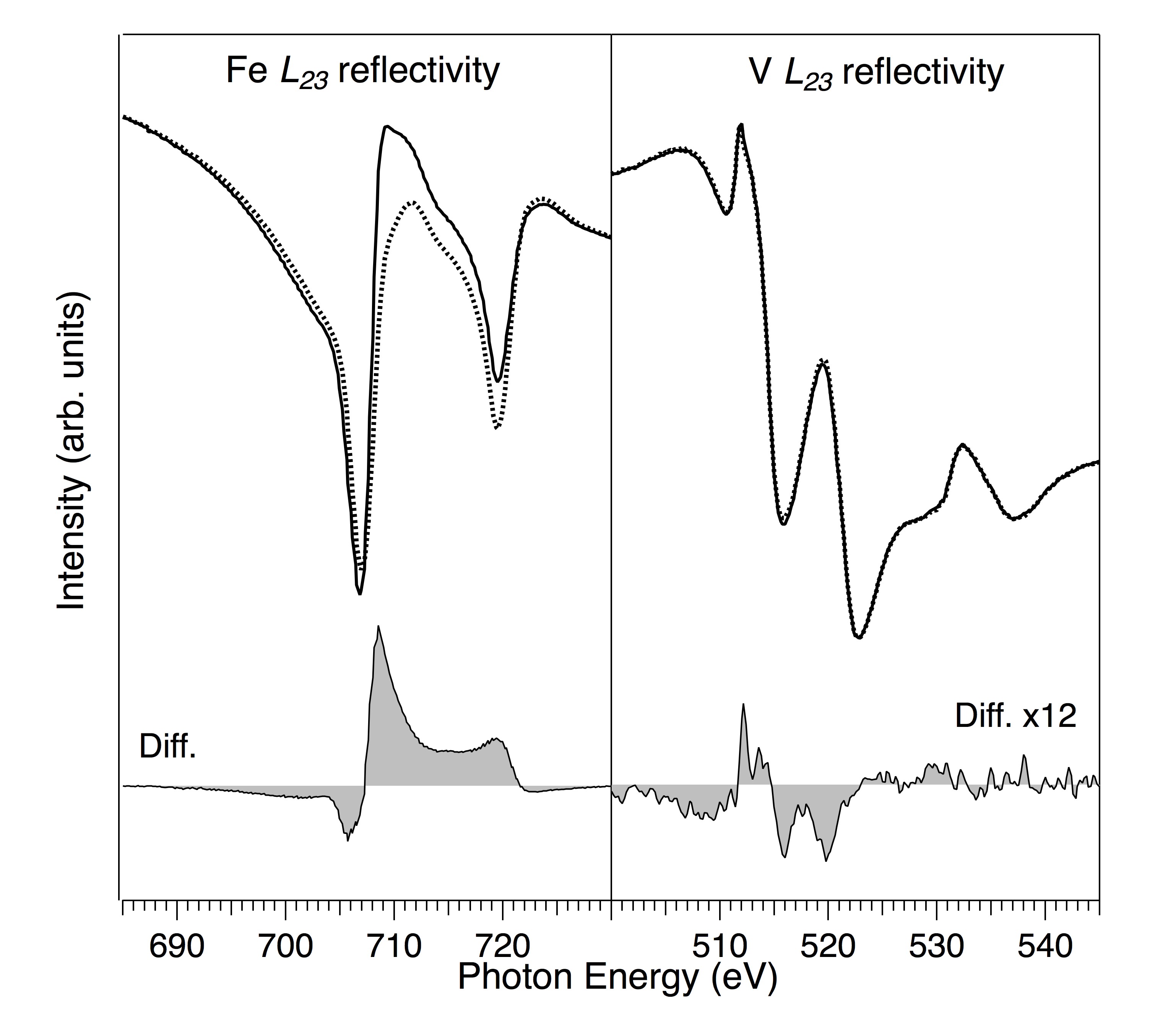}
\vspace{2ex}
\caption{XRMS reflectivity spectra (top) and the corresponding dichroic difference curves (bottom) at the Fe and V $2p$ thresholds for $\theta$ = 10$^{\circ}$. The continuous and dashed curves are the reflectivity spectra for the opposite directions of the applied magnetic field.}
\label{Fig3.jpg}
\end{figure} 

\begin{figure}
\includegraphics[width=85mm]{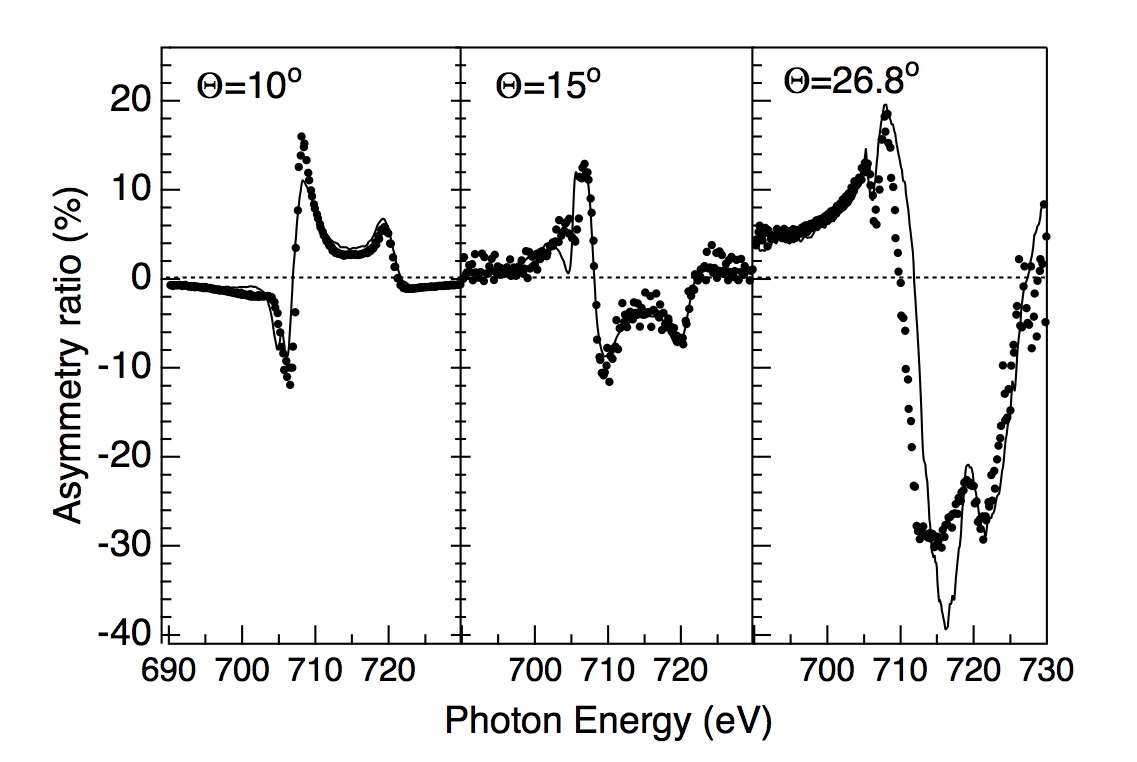}
\vspace{2ex}
\caption{Energy-dependent asymmetry ratios at the Fe $L_{2,3}$ edges for three different scattering angles, including the Bragg angle at 26.8$^{\circ}$. The dots are experimental data and the continuous curves are the results of  simulations.}
\label{Fig4.jpg}
\end{figure}

\section {results and discussion}

Figure 2 shows angle-dependent x-ray diffraction data of the same sample as in Fig. 1 measured at ALS with 704 and 512 eV photon energies at the Fe and V $2p$-edges, respectively. In both cases, the reflectivity is significantly enhanced at low angles and at the Bragg peaks. The first and second order Fe Bragg peaks are observed at 26.8$^{\circ}$ and 63.3$^{\circ}$ and the first order V Bragg peak at 38.6$^{\circ}$. This is consistent with Bragg's law, $\theta_{B}$=arcsin(n$\lambda$)/(2$\times\Lambda$) where, $n$=1 is the first order of diffraction, $\lambda$ is the wavelength and $\Lambda$ is the periodicity (19.5 \AA{}), that gives 26.8$^{\circ}$ and 38.4$^{\circ}$ at the $L_{3}$-edges of Fe (704 eV) and V (512 eV), respectively.
At these angles, the reflected x-rays from all the interfaces in the superlattice interfere constructively and scatter in phase. Note that the intensity depends on the photon energy and  is more than three times larger at the first order Bragg peak of V than for the corresponding Fe peak. The narrow oscillations between the Bragg peaks are Kiessig fringes that are equal to the number of periods in the superlattice. Note that the bandwidths of the Bragg peaks are about 2$^{\circ}$ and correspond to an energy interval of about 20-25 eV with constructive interference.

Figure 3 (top), shows XRMS reflectivity spectra of Fe and V (left and right panels, respectively) at the $2p$ absorption thresholds for opposite directions of the applied magnetic fields (full and dashed curves) at the fixed scattering angle $\theta$=10$^{\circ}$. The peak structures are related to the absorption thresholds with a spin-orbit splitting of 13.1 eV for Fe and 7.7 eV for V as well as the strong interplay between the real and imaginary parts of the resonant magnetic scattering factors at the absorption thresholds. The corresponding dichroic difference spectra are shown below. Both the Fe and V XRMS spectra show magnetization dependence which appears to be mainly antiparallel to each other. While the Fe spectra show a strong dichroic signal, the induced magnetism in V is about 12 times weaker and show more structures. The same type of energy-dependent XRMS spectra were also measured at $\theta$=15$^{\circ}$ and at the Bragg diffraction angles which all show that the induced magnetic moments of V are antiparallel to those of Fe.

\begin{figure}
\includegraphics[width=85mm]{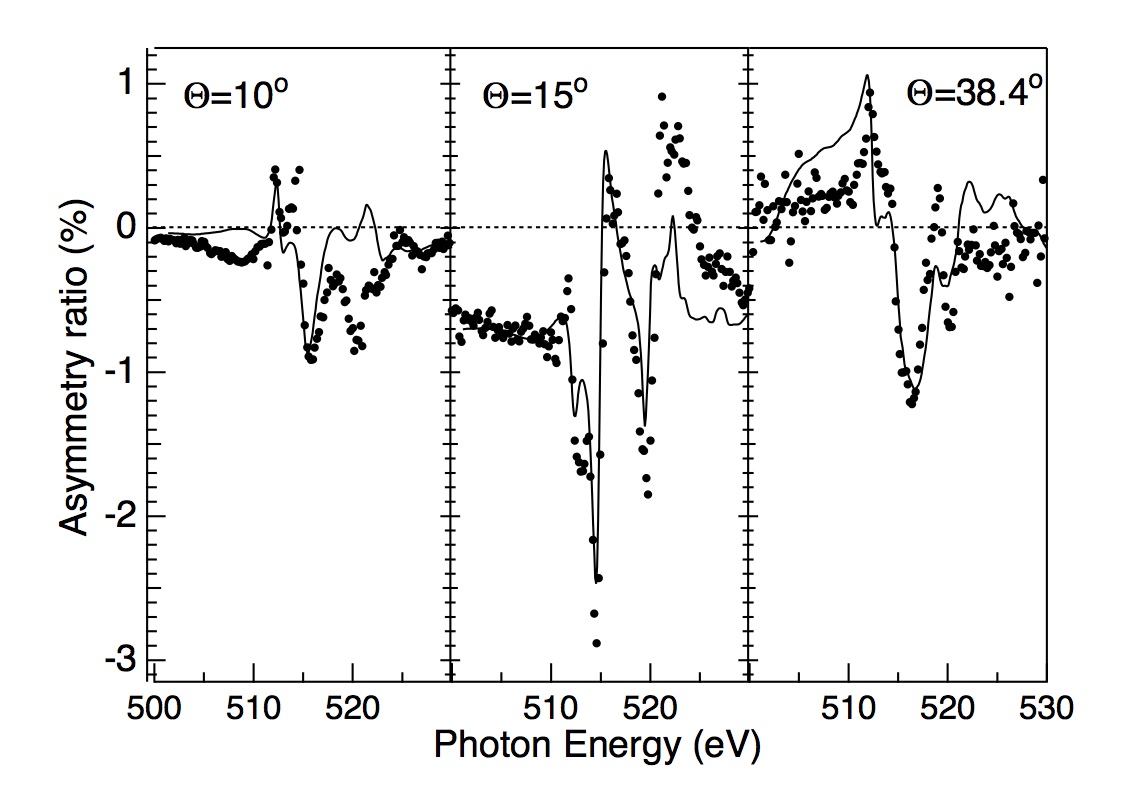}
\vspace{2ex}
\caption{Energy dependent asymmetry ratios at the V $L_{2,3}$ edges for three different scattering angles, including the Bragg angle at 38.4$^{\circ}$. The dots are experimental data and the continuous curves are the results of simulations for the magnetic profile shown in Fig. 4.}
\label{Fig5.jpg}
\end{figure}

The calculations of the energy-dependent x-ray resonant magnetic scattering intensities in the longitudinal mode were made using the program XRMS \cite{XRMS}. This program uses a dynamic optical theory for the x-ray diffraction. The dynamic approach is more complex than a kinematic theory but has the advantage that it can also be applied on resonant magnetic reflectivity measured outside the Bragg peaks. Tabulated values of the real and imaginary parts of the nonresonant anomalous complex scattering factors $f'(E)$ and $f''(E)$ \cite{Henke} of Pd, Fe, V and MgO were used and the average magnetic 
moment of Fe was assumed to be equal to the bulk Fe reference of 2.2 $\mu_{B}$ \cite{Sazaki}. The resonant parts of the $f''(E)$ values corresponding to the absorptive imaginary part of the scattering factors of Fe and V were determined from x-ray absorption measurements \cite{Schwickert} while the $f'(E)$ values corresponding to the dispersive real part were evaluated from the $f''(E)$ data by the Kramers-Kronig relation. The imaginary parts of the magnetic scattering factors $m''(E)$ obtained from the difference curves of XMCD mesurements were scaled in the same way as the $f''(E)$ values. The corresponding dispersive part of the magnetic scattering factor $m'(E)$ was evaluated from the imaginary $m''(E)$ part by using the Kramers-Kronig relation. All reflectivity spectra were fitted with the same structural parameters.

Figure 4 shows energy-dependent asymmetry ratios at the Fe $L_{2,3}$ thresholds for three different scattering angles, including the Bragg diffraction angle at 26.8$^{\circ}$. The dots are experimental data and the continuous curves are results of simulations obtained with the structural parameters given in Table I. The asymmetry ratios are defined as R=(I$^{+}$-I$^{-}$)/(I$^{+}$+I$^{-}$), where I$^{+}$ and I$^{-}$ are the scattered intensities for the two opposite directions of the applied magnetic field. The agreement between the relatively strong measured and simulated asymmetry ratios ($\sim$ 15\%) at the Fe $L_{2,3}$ edges is reasonably good. This confirms that there is ferromagnetic alignment between the different Fe layers. The relatively large variation of the amplitudes of the asymmetry ratios between the different scattering angles is due to the increased probe depth at the Bragg diffraction angle in comparison to the scattering due to reflectivity at the lower angles. Moreover, the magnetic sensitivity at the interfaces is enhanced due to the interference effects at the Bragg angle.

Figure 5 shows energy-dependent asymmetry ratios at the V $L_{2,3}$ thresholds for three different scattering angles, including the Bragg diffraction angle at 38.4$^{\circ}$. The amplitudes of the asymmetry ratios ($\sim$ 1\%) are much weaker than those at the Fe $L_{2,3}$ thresholds and show more complex spectral features with a variation of the amplitudes which largely depend on the scattering angle. This is due to the interference between the real and imaginary parts of the resonant magnetic scattering factors of the different elements involved and the smaller spin-orbit splitting of V in comparison to Fe. The agreement between the measurements and the simulations is reasonably good. Quantitatively, the asymmetry ratios of V are about 12-20 times weaker than for Fe, implying an average induced V magnetic moment in the order of 0.1-0.2 $\mu_{B}$. Due to the interference effects and the enhaced probe depth, the calculated asymmetry ratio at the Bragg angle (38.4$^{\circ}$) was found to be very sensitive to different modeled magnetization profiles while at the other scattering angles dominated by reflectivity, the structural parameters, in particular, the thickness of the capping layer, are more important, as previously shown by e.g., Sacchi \textit{et al.} \cite{Sacchi1}. The magnetic profile was estimated by dividing the V sublayers into slices equal to the regular interatomic distance in the crystalline bct phase of the Fe/V superlattice. For this model, the V $3d$ magnetic profile is determined by refining the energy dependencies of 7 slices, renormalized to unity.

\begin{figure}
\includegraphics[width=80mm]{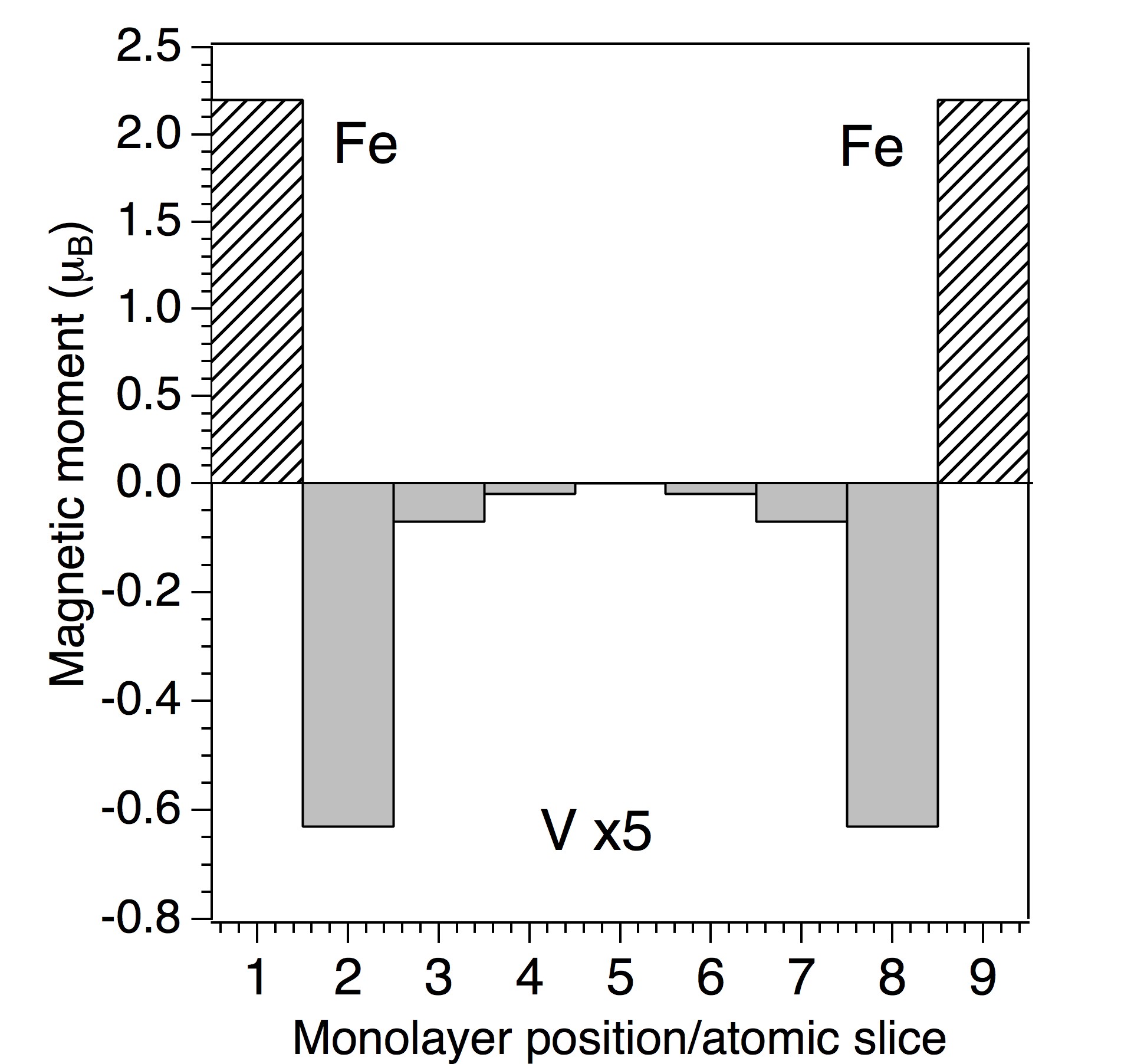}
\vspace{2ex}
\caption{Calculated magnetic profile across the (Fe 6 ML)/(V 7 ML) multilayer. The V moments have been scaled by a factor of 5 to enhance their small values (see text). }
\label{Fig6.jpg}
\end{figure}

Figure 6 shows the magnetic profile estimated from calculations with 7 V monolayers. In the applied model, the number of free parameters was kept as low as possible and we assumed that the magnetic structure is symmetric with respect to the center of the V layer. The imaginary part $m''(E)$ of the mean value of the V magnetic polarization was assumed to be equal to the XMCD amplitude \cite{Schwickert}. Due to the magnetic symmetry constraint of the profile, there are thus three adjustable parameters which are the values of the magnetic polarizations for the first three atomic layers while the values of the other layers all depend on these.

\begin{table}
\caption{Structural parameters for the (Fe 6 ML)/(V 7 ML) superlattice. The total film thickness is 830$\pm$2 \AA{} and $t$=$t_{Fe}$/($t_{Fe}$+$t_{V}$)=0.38. }
\begin{tabular}{ccccc}
\hline
$\textnormal{Period 19.7$\pm$0.1 [\AA{}]} 
$&$\textnormal{Fe}$&$\textnormal{V}$&$\textnormal{Pd}$&$\textnormal{MgO}$\\
\hline
$\textnormal{Thickness [\AA{}]}$&$7.4\pm0.1$&$12.3\pm0.1$&$40.0\pm1.0$&$\infty$\\
$\textnormal{Roughness [\AA{}]}$&$1\pm0.5$&$1.5\pm0.5$&$1.5\pm0.5$&$2\pm0.5$\\
$\textnormal{Density} [10^{3}\textnormal{kg/m}^{3}]$&$7.9$&$6.1$&$12.0$&$1.44$\\
$\textnormal{Atomic weight}$&$55.8$&$50.9$&$106.4$&$20.15$\\
\hline
\end{tabular}
\label{table1}
\end{table}

Due to the large difference in spectral shape between the scattering angles, it was possible to estimate the different parameters even from this relatively limited data set. The description of the spectral shapes of the asymmetry ratios was found to rely on the simulated average interface roughness parameters given in Table I. The refinement procedure with different starting profiles both constant and oscillatory for the three scattering angles always ended up with the exponentially decreasing magnetic profile shown in Fig. 6 where the negative sign of the V polarization has been chosen due to the antiparallel ordering with respect to the Fe magnetization. The amplitudes of the V moments in Fig. 6 have been magnified by a factor of five to highlight their detail. In this model profile, the V atomic polarization in the interface sublayer is 3.5 times larger than the mean value averaged over all V sublayers (-0.18 $\mu_{B}$). It decreases dramatically by a factor of 0.4 in the second atomic layer and by 0.1 in the third atomic layer. The average induced V magnetic moment of about -0.18 $\mu_{B}$ can be used as a scaling factor for an evaluation of the absolute $3d$ magnetic moments in each atomic slice. With this scaling, the magnetic moment of the V atoms directly at the Fe interfaces are strongly polarized with an induced magnetic moment of about -0.63 $\mu_{B}$. For larger distances from the Fe interface, the induced V moments are much weaker, -0.07 $\mu_{B}$ and -0.02 $\mu_{B}$ for the second and third atomic monolayers. 

Previous models for the magnetic profile using standard XMCD in x-ray absorption measurements have indicated that the V moments decay more slowly with distance from the Fe interface and that also the interior of the spacer layers acquires a significant magnetic moment \cite{Tomaz}. It can be anticipated that if XRMS is applied to the same systems, the size of the magnetic moments would decrease much faster \cite{Clavero}. Moreover, the quantitative estimation of the magnetic moments of the early transition metals such as V with XMCD, is based on the validity of the orbital and spin sum-rules that have been debated both experimentally and theoretically \cite{Harp,sumrules3}. In particular, for experimental application of the spin sum rule in XMCD, it is necessary to separate the $L_{3}$ and $L_{2}$ edges which is normally only reliable for systems with a sufficiently large spin-orbit coupling that is only the case for late $3d$ transition metals such as Co and Ni. 

It is noticeable that the obtained average V magnetic moment (-0.18 $\mu_{B}$) is quantitatively in good agreement with XMCD measurements for Fe/V multilayers where an average magnetic moment of about -0.15 $\mu_{B}$ has been determined for 7 vanadium monolayers \cite{Schwickert}. It is also consistent with \textit{ab initio} electronic structure calculations \cite{Niklasson} which indicate that the induced magnetism in the first interface V monolayer oscillates between -0.352 and -0.688 $\mu_{B}$ depending on the lattice constant while the induced magnetism of the inner V monolayers is much weaker. The rapid decrease of the induced magnetic moments in the nonmagnetic spacer layers implies that these layers can be kept relatively thin in the design of a superlattice layer thickness ratio and therefore it may be possible to increase the magnetic coupling strength. These type of studies may therefore contribute to improving devices in applications using the IEC effect such as novel magnetic recording heads, magnetic random-memory cells or magnetic sensors.\\

\section{conclusions}
The induced magnetism of the $3d$ electronic states of vanadium across the individual atomic layers in a (6 ML Fe)/(7 ML V) superlattice has been investigated by bulk sensitive and element selective x-ray resonant magnetic scattering (XRMS). This was done by taking advantage of the enhanced magnetic contrast in the rapidly changing optical properties at the Fe and V $2p$ x-ray absorption resonances in combination with the constructive interference at Bragg scattering conditions. The whole V layer was found to achieve an average induced magnetic moment of -0.18 $\mu_{B}$ that is antiparallel to that of the Fe moments. The results show that the induced magnetic moment of the first V atomic layer directly at the interface with Fe is strongly polarized with a magnetic moment of -0.63 $\mu_{B}$, while the Fe moments remain bulk-like throughout the whole layer. Going from the interface atomic V monolayer next to the magnetic Fe layer into the interior of the V layer, the magnitude of the induced V magnetic moments is more rapidly decreasing than previously observed by standard x-ray magnetic circular dichroism measured in total electron yield mode. The XRMS method together with detailed structural investigations offer a unique possibility to estimate magnetic depth profiles in artificial magnetic structures. Studies of weakly induced magnetic moments in individual atomic layers is useful for gaining detailed insight to the magnetic coupling in a wide range of superlattices as well as, metallic and nonconductive multilayer materials in applied magnetic fields.

\section{acknowledgments}
We would like to thank N. Jaouen for providing the XRMS program, P. Blomqvist for producing the superlattice, M. Sacchi for valuable assistance during measurements and C. Hague for discussions. We are also grateful to E. Gullikson and J.H. Underwood of the Center for X-ray Optics and to the staff at the Advanced Light Source for making these measurements possible. This work was supported by the Swedish Foundation for International Cooperation in Research and Higher Education (STINT).\\

$^{}$Corresponding author: Martin.Magnuson@ifm.liu.se

\end{document}